\documentclass[prb,aps,superscriptaddress,showpacs,floats]{revtex4}
\usepackage{amsmath}
\usepackage{graphicx}
\newcommand{\be}{\begin{equation}}
\newcommand{\ee}{\end{equation}}  
\newcommand{\bea}{\begin{eqnarray}}
\newcommand{\eea}{\end{eqnarray}}
\newcommand{\barr}{\begin{array}}
\newcommand{\earr}{\end{array}}

\newcommand{\ep}{\epsilon}

\begin{document}
\date{\today}

\title{The inelastic Takahashi hard-rod gas}
\author{Umberto~Marini-Bettolo-Marconi} 
\author{Maurizio~Natali} 
\author{Giulio~Costantini}
\affiliation{Dipartimento di Fisica, Universit\`a di Camerino, 
Via Madonna delle Carceri, I-62032, Camerino, Italy}
\author{Fabio~Cecconi}
\affiliation{INFM and Istituto dei Sistemi Complessi (ISC-CNR)\\
Via dei Taurini 19, I-00185, Roma, Italy} 

\begin{abstract}
We study a one-dimensional fluid of hard-rods interacting 
each other via binary inelastic collisions and a short ranged 
square-well potential. 
Upon tuning the depth and the sign of the well, we investigate the interplay 
between dissipation and cohesive or repulsive forces. 
Molecular dynamics simulations of the cooling regime indicate that the 
presence of this simple interparticle interaction is sufficient to 
significantly modify the energy dissipation rates expected by the Haff's 
law for the free cooling.  
The simplicity of the model makes it amenable to an analytical  
approach based on the Boltzmann-Enskog transport equation which allows 
deriving the behaviour of the granular temperature. Furthermore, 
in the elastic limit, the model can be solved exactly to provide
a full thermodynamic description. A meaningful theoretical approximation 
explaining the properties of the inelastic system in interaction with a thermal 
bath can be directly extrapolated from the properties of the corresponding 
elastic system, upon a proper re-definition of the relevant observables. 
Simulation results both in the cooling and driven regime can be fairly 
interpreted according to our theoretical approach and compare rather well 
to our predictions. 
\end{abstract}
\maketitle

\section{Introduction}
Granular materials are ubiquitous in nature and their handling occurs
in many types of industrial activities. While they are very common, 
their properties often are not. 
In the last twenty years there has been
a great progress in the comprehension of static and
dynamical properties of granular 
flows~\cite{general1,general2,general3,general3b,general4}. 
In spite of the fact that most of the theoretical
research in this context
has been based on the inelastic hard sphere model, several
observations suggest that neither cohesive forces~\cite{Schultz}
nor electrostatic repulsion~\cite{Wolf} can be ignored.
Understanding how simple interactions modify the behavior of a granular gas
can have important practical consequences.
Cohesive forces have to be considered when studying wet granular
matter: the humidity may lead to the formation of thin layers
of water on the surface of the grains and induce adhesion through
capillarity effects. The presence of liquid-vapor 
interfaces can enhance the mechanical stability of an assembly of grains,
as illustrated by sand castles~\cite{physicssandcastles}.
On the other side, repulsive forces also play a role, as
stressed  by Sheffler and Wolf~\cite{Wolf}. Dry granular materials tend to 
become electrically charged due to contact electrification during 
transport. In the case of monopolar charging the particles 
experience mutual coulombic repulsion.
Finally,
Blair and Kudrolli~\cite{Blair} studied the behavior of a 
vibrated system of magnetic grains, where 
forces of tensorial character are in action, and found
coexistence of long lived clusters with isolated particles.
Clusters can manifest as chains or globular structures
according to the driving intensity.
 
In this work we introduce and study a one-dimensional model 
which can be tuned to describe both the cohesive and 
the repulsive regime. One dimensional models have 
often been employed in the literature \cite{oned1,oned2,oned3,oned4,oned5} 
to study granular fluids because 
their simplicity provides a valuable 
testing ground for theoretical approaches and approximations. 
Our model consists of a set of inelastic hard rods subjected to 
square-well potential as shown in Fig.~\ref{fig:pot}. 
The attractive potential mimics the action of cohesive forces responsible 
for adhesion among particles which are crucial effects when 
considering fine particulates such as powders or sands.
On the contrary, the barrier describes
the effect of soft materials which may present a 
deformable shell covering the hard-core nucleus. 

The choice of a square well interparticle interaction is particularly
convenient in a computer implementation of the model since it 
reduces Newton's equations to algebraic expressions. Indeed,  
in the cooling regime, rods move with constant velocity until they pass 
a barrier or their cores touch.
Thus the collisional cooling can be 
simulated through the collision driven algorithm of 
Alder-Wainwright~\cite{Alder}.
We shall analyze the interplay between the
potential and the collisional dissipation typical
of granular materials.      
In particular, we discuss the influence of the square-well interaction 
on the rate of energy dissipations in the same spirit
of reference~\cite{Wolf}.
It is well known, that in the homogeneous free cooling process,
a system of inelastic hard spheres dissipates its kinetic energy
at a rate proportional to the square root of the kinetic 
temperature, $T$, the so 
called granular temperature, and that $T$ decreases in time
as the inverse of square time \cite{Haff}. As we
shall see, this picture is partially modified by the presence
of a short range repulsive or attractive potential barrier.
By treating collisions according to Enskog's equation, a generalization
of Boltzmann equation to dense-fluid regime, we are able to make
some predictions about the cooling behavior of the model.    
In addition we consider its properties when it is kept in contact 
to a stochastic source of energy which balances the energy loss
due to inelastic collisions. In this case, the system reaches a steady
regime whose properties can be partly understood through a direct comparison
with the properties of corresponding elastic system.   
      
The paper is organized as follows. 
Section II, describes the model we use and the main 
features of the simulations and technical details. 
Section III shows the thermodynamics of the elastic
version of the model, in order to have a reference 
system to compare inelastic results.
In section IV, an analytic estimate of granular temperature of the system 
is derived through a Boltzmann-Enskog approach.
Section V illustrates simulation results of the 
inelastic model both in the cooling and driven regimes 
with a comparison to theoretical predictions.
Finally section VI contains a brief discussion and conclusions.
     
\section{The Model}
\begin{figure}[htb]
\includegraphics[clip=true,width=8.cm, keepaspectratio,angle=0]
{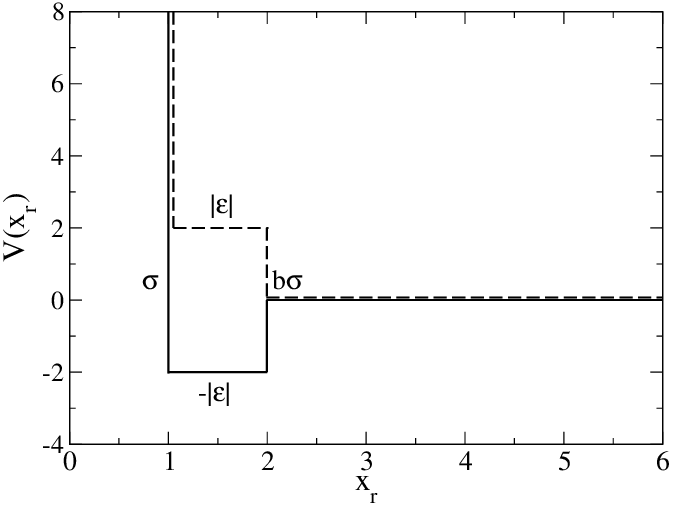}
\caption{Sketch of the interaction potential Eq.(\ref{eq:squarewell})  
as a function of the interparticle distance $x_r$, for $b = 2$ and $\sigma=1$.
Solid line refers to attraction ($\epsilon > 0$) while  dashed to 
repulsion ($\epsilon < 0$).}
\label{fig:pot}
\end{figure}

We consider $N$ identical impenetrable rods of mass $m=1$, size $\sigma=1$,
positions $x_i(t)$ and velocities $v_i(t)$ 
constrained to move in a periodic domain of size $L$.
They interact through a potential $V(|x_i-x_j|)$
consisting of a hard-rod part
and a square-well potential, as shown in Fig.~\ref{fig:pot}.
Explicitly, we consider:
\begin{eqnarray}
V(x) =
\left\{
\begin{array}{ll}
\infty \qquad \qquad \quad \, \, \mbox{if~} x<\sigma\\
-\epsilon \qquad\qquad\quad\mbox{if~}     \sigma\leq x \leq b\sigma\\
0      \qquad\qquad\quad\quad \mbox{if~} x > b\sigma
\end{array}\right.
\label{eq:squarewell} 
\end{eqnarray}
where the parameter $b$ defines the characteristic range of the interaction.
The effect of a piece-wise constant potential $V(x)$ 
amounts to a set of simple collision rules, similar to those involved in the 
dynamics of hard-rods.
Several kinds of collisions between two neighbor 
particles may occur when their distance is:
$$
\Delta x_i(t) \equiv x_{i+1}(t) - x_{i}(t) = b \sigma 
$$
If $\epsilon>0$ the following cases are possible:
I) particles entering the well
II) particles leaving the well, 
III) particles being trapped
and rebounding at the inside square-well edge,
because their relative kinetic energy is not sufficient to escape.

On the other hand,
if $\epsilon<0$ one has the cases:
I') particles overcoming the repulsive barrier ($m(v_i-v_{i+1})^2>4|\epsilon|$)
II') particles descending the barrier,
III') particles being repelled by the barrier, when $m(v_i-v_{i+1})^2<4
|\epsilon|$.

The post-collisional velocities in cases III and III' are given by:
\begin{eqnarray*}
& v_i'     & = v_{i+1} \\
& v_{i+1}' & = v_i.
\end{eqnarray*}

In the remaining cases the collision rule are found
by requiring again the conservation of the total energy and total 
momentum at 
the edge of the square well-potential. If particles are entering
($s_i \equiv \mbox{sign}(v_i-v_{i+1})>0$), while if they 
are leaving the well ($s_i<0$),
and the collision rule reads:
\begin{equation}
\begin{aligned}
v'_{i} = \frac{(v_i+v_{i+1})}{2}
         +  s_{i}\sqrt{\frac{(v_i-v_{i+1})^2}{4} 
         +  s_{i}\frac{\epsilon}{m}}\quad \\
v'_{i+1} = \frac{(v_i+v_{i+1})}{2} 
         -  s_{i}\sqrt{\frac{(v_i-v_{i+1})^2}{4} 
         +  s_{i}\frac{\epsilon}{m}}
\end{aligned}
\label{v1-v2}
\end{equation}
where pre-collisional 
and post-collisional velocities 
are indicated by unprimed 
and primed symbols, respectively.

We finally consider the hard-core inelastic collision at
$$
\Delta x_i(t) \equiv x_{i+1}(t) - x_{i}(t) = \sigma
$$
which results in the transformation
\begin{equation}
\begin{aligned}
v_{i+1}' = v_{i+1}-\frac{1+\alpha}{2}(v_{i+1}-v_i)\\
v_{i}' = v_{i}+\frac{1+\alpha}{2}(v_{i+1}-v_i)\quad\quad
\end{aligned}
\label{collision}
\end{equation}
where $\alpha$ indicates a constant coefficient of restitution
and $0\leq \alpha\leq 1$.

Besides impulsive forces between particles, we shall also consider
an external stochastic
white noise force, whose role is to fluidize the system
and balance the energy losses due to dissipative forces.
The dynamics between two consecutive collisions is described
by the following Langevin equation 
\begin{equation}
m\frac{ d^2 x_i(t)}{ d t^2}= -m\gamma\frac{ d x_i(t)}{ d t }+ 
{\xi_i},
\label{eq:due}
\end{equation}
where $-m\gamma d x_i/dt$ is a viscous term and $\xi_i$ a
Gaussian random force, with zero average and variance 
satisfying a fluctuation-dissipation relation:
\begin{equation}
\langle \xi_i(t) \xi_j(t') \rangle = 2 m \gamma
T\delta_{ij} \delta(t-t')
\label{variance}
\end{equation}
with $T$ proportional to the intensity of the stochastic 
driving~\cite{puglisi,puglisi1}. 
The damping term renders the 
system stationary even in the absence of collisional dissipation and
physically can represent the friction between the particles and the
container.
Summarizing, the position $x_i$ ($i=1,N$) of the $i-$th particle evolves
according to the equation:
\begin{equation}
m \frac{d^2 x_i(t)}{dt^2} =
-m \gamma\frac{d x_i(t)}{dt} + \xi_i(t) + \sum_j f_{ij}(t)
\label{kramers}
\end{equation}
where $f_{ij}$ indicates the resultant of impulsive forces 
between particles $i$ and $j$. 
Since the dynamics of the model is mainly ruled by impulsive forces, 
Molecular Dynamics (MD) simulations make use of a collision
driven algorithm.

\section{Elastic system: equilibrium properties}
The elastic fluid, corresponding to the limit $\alpha=1$ in 
Eqs.~(\ref{collision}), 
serves conveniently as a reference system. Thus, we consider its 
equilibrium properties, that we shall compare to properties of the 
stationary inelastic system to build a theoretical approach valid in the
region of moderate inelasticity.
The equilibrium square-well fluid model is exactly solvable
when the interaction range is 
restricted to first neighbors, 
\textit {i.e.} $b \leq 2$, the exclude volume allows no more than two 
particles to experience the same potential well. 
In that case,  
the Gibbs free energy, $G(P,T,N)$, can be derived using the 
isothermal-isobaric ensemble~\cite{Takahashi,Bishop}.
Here, the partition function $Y(P,T,N)$ is related to that of 
the one-dimensional 
canonical ensemble $Z(T,L,N)$ by
\begin{equation}
Y(P,T,N) =\frac{1}{\Lambda^{N} \Lambda_0}\int_0^\infty  dL
{\mbox e}^{-\beta PL} Z(T,L,N) 
\end{equation}
where $P$ is the thermodynamic pressure, 
$\Lambda\equiv h/\sqrt{2 \pi m k_B T}$ 
is the temperature-dependent de Broglie wavelength and $\Lambda_0$ an 
arbitrary constant with dimension of a length.

Following the existing  literature~\cite{Bishop}, 
the isothermal-isobaric partition 
function for $N$ rods of length $\sigma$ can be written as
\begin{equation}
Y(P,T,N) =\frac{\Lambda}{\Lambda_0}
\Big\{\frac{1}{\beta P\Lambda}\Big[\mbox{e}^{\beta \ep}
(\mbox{e}^{-\beta P \sigma}-
{\mbox e}^{-\beta P b \sigma}) +\mbox{e}^{-\beta P b \sigma }\Big]\Big\}^{N+1}.
\label{Yesplicit}
\end{equation}
 
The associated Gibbs potential 
$$
G(P,T,N)=-\frac{1}{\beta}\ln Y(P,T,N)
$$
reads a part from a constant
\begin{eqnarray}
G(P,T,N)=
(N+1)\Big\{b\sigma P \nonumber 
+\frac{1}{\beta}\ln\Big(\beta P \Lambda\Big)\\
-\frac{1}{\beta}\ln[1+e^{\beta \ep}(e^{\beta P (b-1)\sigma}-1)]\Big\}.
\end{eqnarray}
The equation of state, relating density, pressure and 
temperature is obtained by 
differentiating $G$ with respect to $P$ and defining the 
``volume'' per particle, $\rho ^{-1}=\bar{L}/N$, of the system 
\begin{equation}
\frac{1}{\rho}= b\sigma + \frac{1}{\beta P}
-\frac{(b-1)\sigma}{1 + \mbox{e}^{-\beta P (b-1)\sigma}(
\mbox{e}^{-\beta \ep} -1)}.
\label{rhom1}
\end{equation}
that can be recast  to the more familiar form
\begin{equation}
\frac{\beta P}{\rho} = \frac{1}{1 - \rho \sigma B}
\label{eq:state}
\end{equation}
with 
$$
B = \frac{1 + b \mbox{e}^{-\beta P (b-1)\sigma}(
\mbox{e}^{-\beta \ep} -1)}{1 + \mbox{e}^{-\beta P (b-1)\sigma}(
\mbox{e}^{-\beta \ep} -1)}
$$
Notice that $b=1$ implies $B\to 1$, therefore the hard-rod pressure is 
straightforwardly recovered. 

In order to apply Enskog's kinetic approach,
we need to compute the
equilibrium pair correlation function in the thermodynamic
limit. 
The pair correlation is defined as: 
\begin{equation}
\rho g(y)\equiv\sum_{r=1}^{\infty}\langle \delta(x_{l+r}-x_l-y)\rangle 
\label{ave}
\end{equation}
To perform the average in eq.~(\ref{ave}), we 
represent the delta distribution (with $l=1$ and $q=r+l$) by its Fourier
transform
\begin{equation}
\delta(x_{q}-x_1-y)\equiv \int^{\infty}_{-\infty}\frac{dk}{2\pi}\exp[ik(x_{q}-x_1-y)]
\end{equation} 
and write the average explicitly in terms of the Boltzmann weights
$f(x)=\exp[-\beta V(x)]$:
\begin{eqnarray}
\langle\delta(x_{q}-x_1-y)\rangle = \frac{1}{Z(T,L,N)}\int^{\infty}_{-\infty}\frac{dk}{2\pi} e^{-iky}\int^{L}_{0}dx_{q}\int^{x_{q}}_{0}dx_{q-1} \cdot\cdot\cdot\int^{x_{2}}_{0}dx_{1}\nonumber \qquad \qquad \\
\cdot f(L-x_{q})f(x_{q}-x_{q-1})\cdot\cdot\cdot 
f(x_2-x_1)f(x_1)e^{ik(x_{q}-x_1)} \qquad\qquad \qquad . 
\end{eqnarray} 
Since the last expression has the form of an iterated convolution,
one can obtain the desired average by means of standard 
Laplace transform method, a simple generalization
of the method~\cite{Bishop} employed to compute $Z(T,L,N)$ 
in the isothermal-isobaric ensemble.
After a lengthy calculation, in the thermodynamic limit
$N\to\infty$ and constant pressure, we get the series
\begin{equation}
\rho g(y)=\sum_{r=1}^{\infty}
\frac{A_r(y)}{(r-1)!}\frac{(\beta P)^r e^{-\beta P y}}{[e^{\beta \ep}
(e^{-\beta P \sigma }-
e^{-\beta P b \sigma})
+e^{-\beta P b \sigma }]^{r}}
\label{eq:rhog}
\end{equation}
where the coefficients can be written as
\begin{eqnarray}
A_r(y) = \sum_{k=0}^r
{r \choose k}
\Theta [y-\sigma (r+(b-1)k)] \nonumber \\
\cdot [y- \sigma(r+(b-1)k)]^{r-1}(1-e^{\beta \ep})^k 
(e^{\beta \ep})^{r-k}
\label{coeffic}
\end{eqnarray}
where $\Theta(x)$ is the unitary step function~\cite{note}.
Notice that for a pure hard-rod system ($\ep=0$, 
$\beta P=\rho/(1-\rho\sigma)$), one finds
\begin{eqnarray}
\rho g_{hr}(y)=
\sum_{r=1}^{\infty}\frac{1}{(r-1)!}\frac{\rho^r}{(1-\rho\sigma)^r}
\Theta (y-r\sigma)(y-r\sigma)^{r-1} \nonumber \\
\cdot \exp \Big[ -\frac{\rho}{1-\rho\sigma}(y-r\sigma) \Big ] \qquad \qquad \qquad \quad
\end{eqnarray}
a result which agrees with the well known result by Zernike and 
Prins~\cite{Zernike,Salsburg}. 
From expressions~(\ref{eq:rhog}) and (\ref{coeffic}), 
the value of the pair correlation at contact can be extracted
explicitly: 
\begin{equation}
g(\sigma) = \frac{\beta P}{\rho} 
\frac{1}{1 +  \mbox{e}^{-\beta P(b-1)\sigma} 
(\mbox{e}^{-\beta \epsilon} - 1)}
\label{gsigma}
\end{equation}
Since the thermodynamic pressure
can be expressed in terms of $g(x)$
through the virial equation \cite{virialST}
\begin{equation}
\frac{\beta P}{\rho}=1-\frac{\beta\rho}{2}
\int_{-\infty}^{\infty} dx V'(x)x g(x)
\label{viriale}
\end{equation}
after some rearrangements, the pressure reads
\begin{equation}
\frac{\beta P}{\rho}=1 + \rho\sigma[g(\sigma) + b g(b\sigma^{+}) - 
b g(b\sigma^{-})]
\label{veroviriale}
\end{equation}
where the discontinuity of the potential (\ref{eq:squarewell}) 
at $x=b\sigma$ results in a jump of the binary correlation function:
\begin{eqnarray}
g(b\sigma^{+}) & = & g(\sigma) e^{-\beta P(b-1) \sigma - \beta\epsilon}
\label{gpiu} \\
g(b\sigma^{-}) & = & g(\sigma) e^{-\beta P(b-1) \sigma} \label{gmeno}  
\end{eqnarray}
with
$g(b\sigma^{\pm})=\lim_{\delta\to 0}g(b\sigma\pm\delta)$.
These relations are consistent with Eq.~(\ref{rhom1}) and show that 
the pressure depends not only on the value of the pair correlation at contact, 
but also on its values at $x=b\sigma$.

\section{Boltzmann-Enskog equation}
The system dynamics is determined by the combined effects of
the heat-bath and interparticle collisions. Thus, the 
one-particle phase distribution function $f(x,v,t)$, in the absence  
of external drift and large density fluctuations, evolves 
under the action of a Kramers operator associated with the 
interaction with the heat-bath 
$$
{\cal L}_{K}=\frac{\gamma}{m\beta}\frac{\partial^2 }{\partial v^2}
+\gamma\frac{\partial}{\partial v}v 
$$
plus a collision 
operator, that we represent for the sake of simplicity 
as a Boltzmann-Enskog collision integral:
\begin{equation}
\frac{\partial f(v,t)}{\partial t}={\cal L}_{K}f(v,t)+I(v,t).
\label{evo}
\end{equation}
Adapting to the present case arguments \cite{Stell,Russi} 
similar to those leading to the derivation of the
SET (Standard Enskog Theory)  we
arrive at the following form of 
the Boltzmann-Enskog collision integral $I(v,t)$:
\begin{equation}
I(v,t) = I_0(v,t) + I_{+}(v,t) + I_{-}(v,t) + I_{BS}(v,t)
\label{bolt}
\end{equation}
where the four contributions represent, respectively:

\noindent 
$\bullet$ the inelastic hard-core collision 
\begin{eqnarray}
I_0(v,t) =g(\sigma) \int dv'\int dv''
|v'-v''|\nonumber \\
\cdot f(v')f(v'')\{\delta[v-\frac{1-\alpha}{2}v'-\frac{1+\alpha}{2}v'']
-\delta[v-v'']\}\nonumber \\
\label{I1}
\end{eqnarray}
\noindent
$\bullet$ the entering collision ($(v'-v'')>0$)
\begin{eqnarray}
I_{+}(v,t) =g(b\sigma^+)\int dv'\int dv''
\Theta[(v'-v'')^2+\frac{4\epsilon}{m}]
\nonumber \\
\cdot
\Big\{\delta\Big[v-\frac{v'+v''}{2}+\sqrt{\frac{(v'-v'')^2}{4}+
\frac{\epsilon}{m}}\Big]  \nonumber  \\
-\delta[v-v'']\Big\}|v'-v''|
f(v')f(v'')\nonumber \\ 
\label{I2}
\end{eqnarray}
\noindent
$\bullet$ the escape collision ($(v'-v'')<0$)
\begin{eqnarray}
I_{-}(v,t) =g(b\sigma^-)\int dv'\int dv''
\Theta[(v'-v'')^2-\frac{4\epsilon}{m}]\nonumber\\
\cdot
\Big\{\delta\Big[v-\frac{v'+v''}{2}-\sqrt{\frac{(v'-v'')^2}{4}-
\frac{\epsilon}{m}}\Big] \nonumber \\
-\delta[v-v'']\Big\}|v'-v''|f(v')f(v'')\nonumber \\
\label{I3}
\end{eqnarray}
\noindent
$\bullet$ the elastic bound-state collision at $\Delta x=b\sigma^{-}$
$$
I_{BS}(v,t) = 0
$$ 
which in one dimension vanishes and therefore can be omitted.

We can apply the previous analysis to the theoretical description 
of the cooling process in the presence of the interparticle potential, 
under the hypothesis of spatial homogeneity. 
By integrating with respect to $v$ the second term in the right
hand side of Eq.~(\ref{I1}-\ref{I3}) and approximating $f(v)$ with 
Maxwellian distribution of temperature $T_g$, we obtain the Enskog collision 
frequency at $\Delta x=\sigma$, 
and the frequencies of entering and escaping collisions:
\begin{eqnarray}
\begin{array}{ll}
\omega_{0}(\rho\sigma)=\langle |v_{rel}| \rangle \rho g(\sigma) \qquad \quad \\
\\ 
\omega_{+}(\rho\sigma)=\langle |v_{rel}| \rangle \rho g(b\sigma^{+})
[\Theta(\epsilon)+\Theta(-\epsilon)e^{\beta\epsilon}]
\qquad \quad \\
\\
\omega_{-}(\rho\sigma)=\langle |v_{rel}| \rangle \rho g(b\sigma^{-})
[\Theta(-\epsilon)+\Theta(\epsilon)e^{-\beta\epsilon}]
\end{array}
\label{eq:frequencies} 
\end{eqnarray}
The expression for $\omega_0$ is formally identical to that 
obtained in the case of simple hard-rods without potential tail.
It can be easily verified that the two factors containing the 
$\Theta$-functions are exactly the terms that compensate 
the asymmetry coming from expressions~(\ref{gpiu}) and 
(\ref{gmeno}). 
Therefore, the rates become equal
\begin{eqnarray}
\omega_{+} = \omega_{-} =
\left\{
\begin{array}{ll}
\langle |v_{rel}|\rangle\rho g(\sigma)e^{-\beta P(b-1)\sigma}  \quad\quad\quad \mbox{if~} \epsilon<0\\
\langle |v_{rel}|\rangle\rho g(\sigma)e^{-\beta P(b-1)\sigma-\beta \epsilon}  \quad \quad \mbox{if~} \epsilon>0
\end{array}\right.
\label{eq:rates}
\end{eqnarray}
and thus satisfy a detailed balance relation between entering and escape 
collisions. 
The presence of the potential is reflected in the modified
value of the pair correlation at contact Eq.~(\ref{gsigma}).
By substituting $\langle |v_{rel}| \rangle = 2(\beta\pi m)^{-1/2}$
we find the following expression for the 
collision time of the square well fluid:
\begin{equation}
\omega_0= 2 \sqrt{\frac{\beta}{\pi m}}
\frac{P}{1 + e^{-\beta P(b-1) \sigma}(e^{-\beta \ep}-1)}
\label{eq:omega0}
\end{equation}
For the sake of comparison the hard-rod Enskog frequency reads:  
\begin{equation}
\omega_{hr}= 2 \sqrt{\frac{\beta}{\pi m}}P_{hr}.
\end{equation}
with $\beta P_{hr}=\rho/(1-\rho \sigma)$.
 
Interestingly,
in an elastic system with repulsive forces, the ratio between
the hard-core collision frequency and the entering/escape frequency 
reads
\begin{equation}
\frac{\omega_{0}}{\omega_{\pm}} = e^{\beta P(b-1) \sigma}\;,
\label{eq:omegab}
\end{equation}
it suggests that, at high densities
hard-core collisions dominate, because the pressure is a growing function of 
the density.  Therefore, increasing the density amounts somehow to lowering 
the height of the effective potential barrier, {\it i.e.} the kinetic energy 
required to perform an elastic collision.

We now consider, how the average kinetic energy of the inelastic
system ($\alpha<1$) is dissipated.
By multiplying Eqs. (\ref{bolt})
by $v^2$ and integrating over the velocity, we can compute the
loss of kinetic energy due to collisions.
Since only hard-core collisions
dissipate energy we find that solely the process represented by
Eq.~(\ref{I1}) contributes to the 
evolution equation~\cite{Ernst} for the granular
temperature $T_g(t)$,
\begin{equation}
\frac{{\rm d}T_g}{{\rm d}t}= -(1-\alpha^2)\omega_0 T_g\;.
\label{eq:Trateb}
\end{equation}
 
Notice that the expression (\ref{eq:omega0}) for $\omega_0$, entering 
equation~(\ref{eq:Trateb}), employs a pair 
correlation function $g(\sigma)$ extrapolated from its 
equilibrium value, where $\beta$ has been identified with 
the inverse granular temperature, {\em i.e.} $\beta=1/T_g$.
Moreover, the value of the pressure
necessary to compute the frequency $\omega_0$ can be obtained numerically 
by inverting Eq.~(\ref{rhom1}) for a given density of the system. 
The rate $\omega_0$ decreases with increasing
the repulsive barrier ($\epsilon \to -\infty$)
or when the temperature tends to zero. 
Consequently, as
the system cools down, the dissipation rate will be much slower
than the corresponding rate when $\epsilon=0$.

\section{Numerical results}
Whereas the equilibrium properties of the conservative system 
are analytically accessible, most of the properties of its dissipative 
version need MD simulations to be investigated. After resorting to 
numerical methods we shall compare their results with our theoretical
estimates. 
At values of the restitution parameter $\alpha$ less than $1$,
the system is certainly not in thermodynamic equilibrium, 
but can achieve a stationary state when in contact with the heat-bath
described by Eq.~(\ref{variance}).

We shall consider the behavior of the system both in the cooling regime
($\gamma=0$ and $T=0$) and in the stationary heated regime.  
The numerical methods employed have been briefly mentioned in the 
previous section and described in detail in 
papers~\cite{dererum1,dererum2} which the reader can refer to.

In order to minimize surface effects and simulate an infinite system,
periodic boundary conditions are imposed on the equations of motion. 
During each simulation run, we monitor the kinetic temperature, named
granular temperature, $T_g$, which by definition, 
is proportional to the average of the kinetic energy per particle:
\begin{equation}
T_g = \frac{1}{N}\sum_i^N m [\langle v_i^2 \rangle - 
                             \langle v_i \rangle ^2 ]
\label{Tg}
\end{equation}
having chosen units in which $k_B=1$.
The values of pressure, instead, can be obtained from the virial formula 
\cite{virialMD} properly modified for the present system:
\begin{equation}
\frac{PL}{N T_g}=1+\frac{1+\alpha}{2}\frac{m}{t_{ob} N T_g } Z
\label{Pg}
\end{equation}
where $Z$ indicates the sum  
\begin{equation}
Z = \sum_{all~coll} x_{ij} v_{ij}
\end{equation}
$t_{ob}$ is the observation time and 
$x_{ij}=(x_i-x_j)$ and $v_{ij}=(v_i-v_j)$ are, respectively, the separation 
and the relative velocity at the moment of collision. 
Both kinds of collisional events $|x_{ij}| = \sigma$ 
and $|x_{ij}| = b \sigma$ determine an exchange of momenta
among the particles.

\subsection{Cooling regime}
We consider, first, the properties of a system of $N=2000$ particles 
evolving without the presence of heat-bath, thus no energy injection ($T=0$) 
and no 
friction ($\gamma=0$) occur. In the literature this situation is 
generally referred to as free cooling. 
The properties of this system with $\ep=0$ have been studied
thoroughly~\cite{oned1,oned2,oned3,oned4,oned5} and are well known.
Due to the repeated inelastic collisions, 
the temperature $T_g$ decreases and after a 
short transient, lasting only few collisions per particle,   
$T_g$ displays the typical power law behavior $t^{-2}$, 
known as Haff's law.
During such a regime, the density remains homogeneous
and the velocity distribution converges, from the initial Maxwellian,
to a two hump function.
As the system cools down, particles cluster into two ``streams'' 
at the outer edges the distribution and a bimodal velocity distribution 
emerges. 

\begin{figure}[htb]
\includegraphics[clip=true,width=8.cm,keepaspectratio,angle=0]
{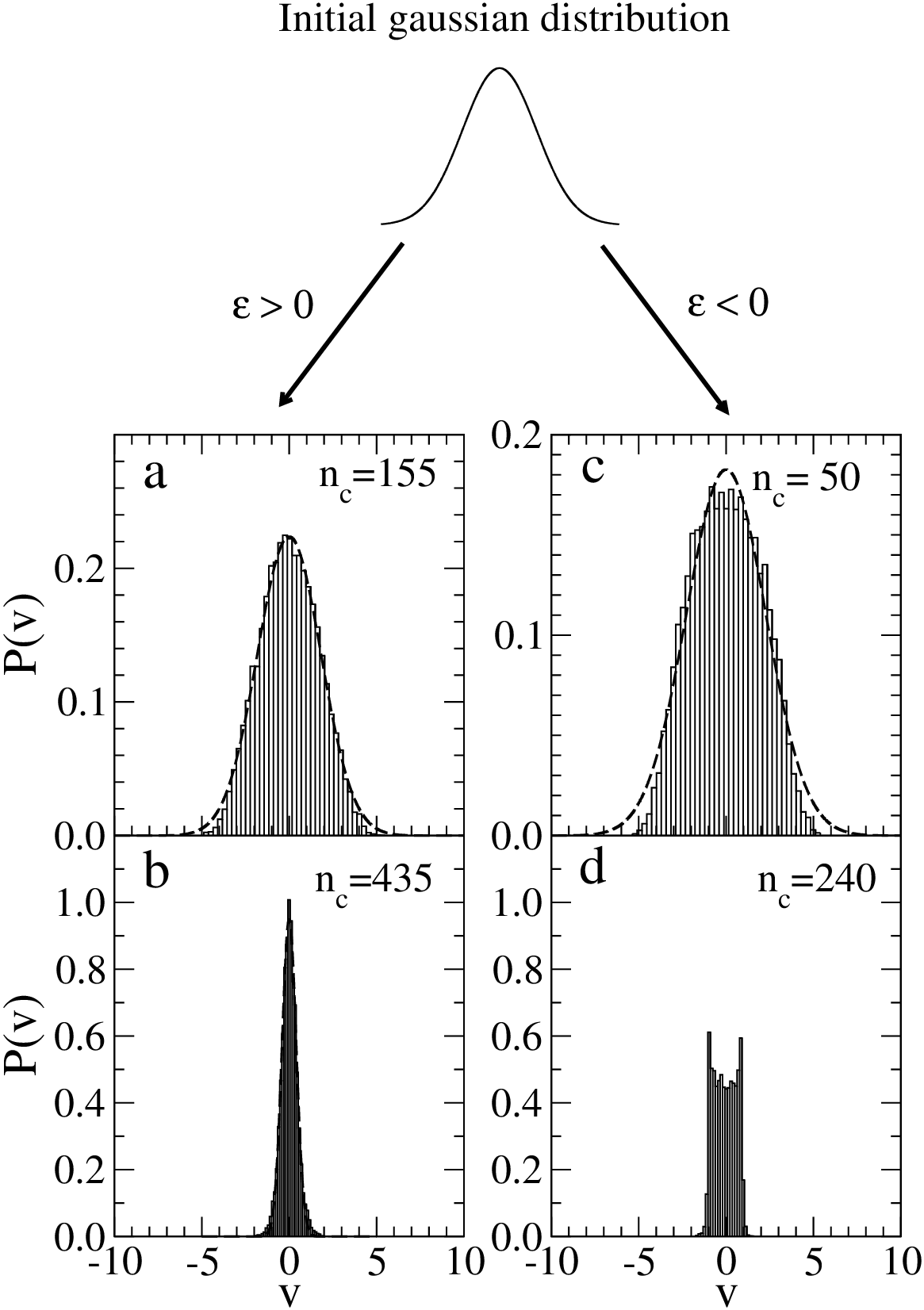}
\caption{Quenching of particle velocities observed at two stages of the
cooling process of a system with attractive ($\epsilon=1$, left) 
and repulsive ($\epsilon=-1$, right) interparticle interaction. 
The histograms of the rescaled (dimensionless) velocity $u=v \sqrt{m/|\ep|}$ 
are collected after $n_c$ hard-core collisions per particle have occurred. 
Simulations refer to a system of $N=2000$ particles and
parameters  
$\alpha=0.99$, $\rho\sigma=0.002$ and $b=2$. The  
the Gaussian fits (dashed lines) are plotted for comparison.}
\label{fig:dist}
\end{figure}

Our MD simulations show that this scenario is modified by the
the presence of potential tail (see Eq.~(\ref{eq:squarewell})). 
Every MD run starts from an initial state characterized by $N=2000$ 
particles with 
a Maxwellian velocity distribution of temperature $T$ and uniformly 
distributed in space with no overlaps. 
During the dynamics, the grains spontaneously organize toward a state 
where the velocity distribution $P(v)$ depends on the attractive or repulsive 
character of $V(x)$. The behaviour of $P(v)$ is clearly shown in 
figure~\ref{fig:dist}, where later time distributions are characterized by a 
nearly Gaussian shape for $\ep>0$ and no longer Gaussian for $\ep<0$.
The attractive interaction has the effect to accelerate the 
dissipation,
however the velocity distribution does not display the typical two-hump 
feature proper of the $\epsilon=0$ case, remaining a single peak function of
shrinking width (see Fig.\ref{fig:dist} left panel).    

On the contrary, when the potential is repulsive, only those pairs 
with velocities satisfying the condition $(v_{i} - v_{i+1})^2 > 4|\epsilon|/m$
may perform inelastic collisions. Such a selection mechanism is irrelevant 
when $T_g >> \epsilon$, {\em i.e.}
the very early stages of the simulations, however, it 
eventually leads to velocity distributions with small tails outside the region  
$-\sqrt{|\ep|/m}\leq  v  \leq \sqrt{|\ep|/m}$ and almost flat 
inside (Fig.~\ref{fig:dist} right panel). 
Two small peaks can also be observed at $v = \pm \sqrt{|\ep|/m}$, likely, 
a reminiscence of the free-cooling two stream mechanism.

\begin{figure}[htb]
\includegraphics[clip=true,width=8.cm, keepaspectratio,angle=0]
{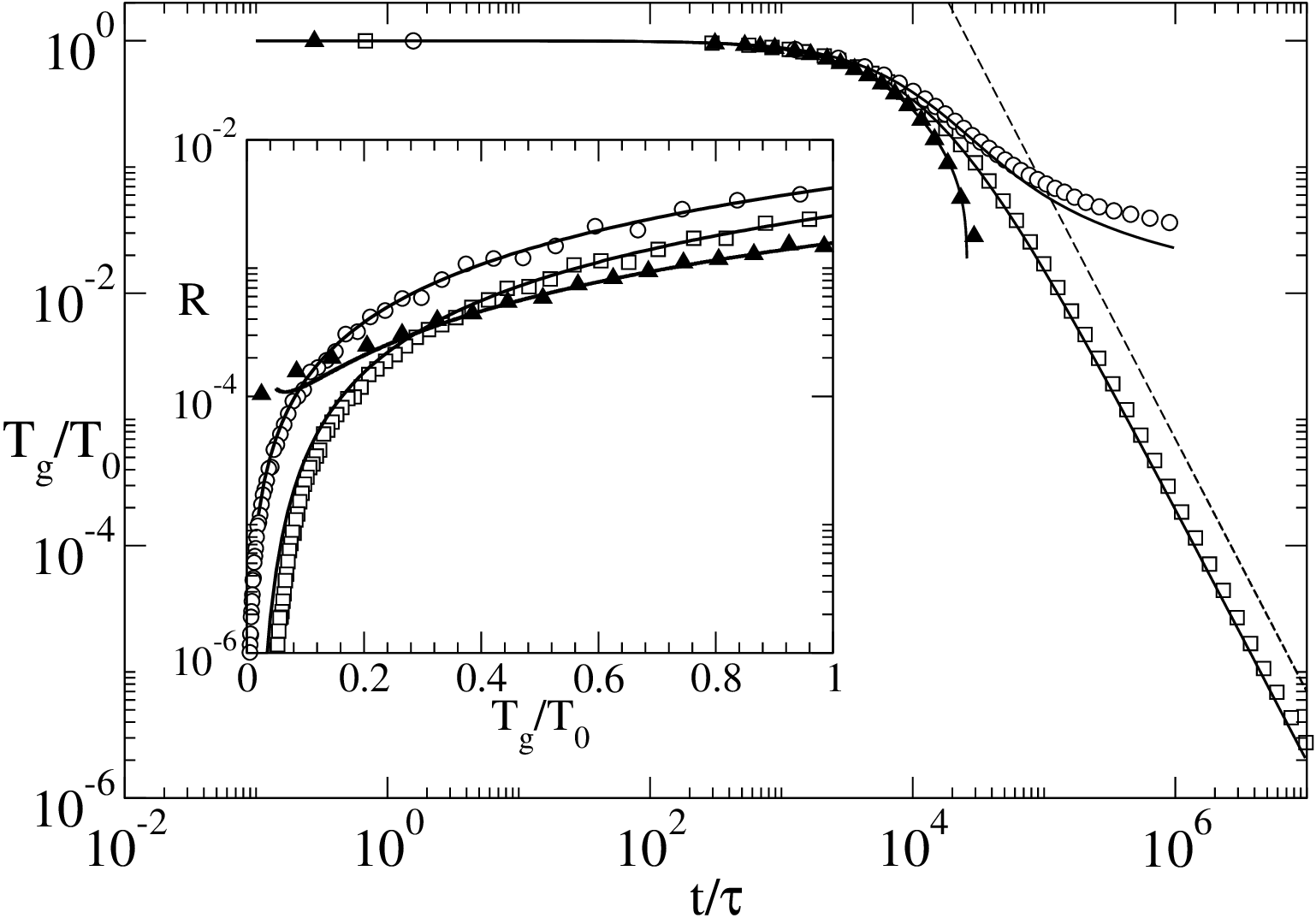}
\caption{Simulation results of the energy decay 
with time measured units $\tau = \sigma\sqrt{m/T_0}$,  
for $\epsilon=0$ (squares),
$\epsilon=-1$
(circles) and  $\epsilon=1$ (triangles) at effective density 
$\rho\sigma=0.002$.  Each point is the average over one-hundred  
trajectories of a system with $N=2000$ hard rods and 
initial temperature $T_0 = 10$.
Lines represent the analytical estimate from Eqs.~(\ref{eq:Trateb}) 
and (\ref{eq:omega0}), coherently with Eq.~(\ref{rhom1}).
Inset: Plot of the theoretical and numerical (dimensionless) dissipation 
rate $R = \tau \dot{T}$ versus the rescaled granular temperature
(same symbols).}
\label{fig:energydecay}
\end{figure}
Equation (\ref{eq:omega0}) indicates that, under repulsive  
interaction, particles collide inelastically with an initial rate  
$\omega_0\propto \sqrt{T_g}$, that,  
as the system cools down, makes the crossover to  
the behavior $\omega_0 \propto \sqrt{T_g}\exp(\ep/T_g)$.
Accordingly, fewer and fewer particle 
pairs will collide and the cooling slows down leading to a logarithmic
decay in time of the temperature. However, this argument turns out to
be incorrect. 
Indeed, Fig.~\ref{fig:energydecay} proves that 
the energy dissipation process occurs with a slower time decay than the  
prediction given by Eq.~(\ref{eq:Trateb}). The direct comparison between
theoretical and simulated dimensionless rate $R= \tau \dot{T_g}/T_0$ 
is shown in the inset, where $\tau$ is a proper time scale. 
The reason for such a discrepancy 
relies on the fact that the Maxwellian approximation for the velocity 
distributions, used to derive the rate expression (\ref{eq:omega0}), fails 
as it seen from Fig.~\ref{fig:dist} (right). 
With the actual shape of the distribution $P(v)$, indeed,    
the system dissipates only a negligible fraction of the kinetic
energy and undergoes an effective re-elasticization, implying that
the non-Maxwellian character of $P(v)$ is maintained up to the inelastic 
collapse. Our theoretical estimate  
of the collision frequencies works better at moderate densities, where 
dissipation can counterbalance the re-elasticization. 
As it suggested by Eq.~($\ref{eq:omegab}$) indeed,
The pressure exerted by the dense surrounding fluid on
two colliding partners may overwhelm
their repulsion, so that they will experience frequent hard-core 
collisions, {\em i.e.} $\omega_0 >> omega_{\pm}$.




\subsection{Driven regime}
\begin{figure}[htb]
\includegraphics[clip=true,width=8.cm, keepaspectratio,angle=0]
{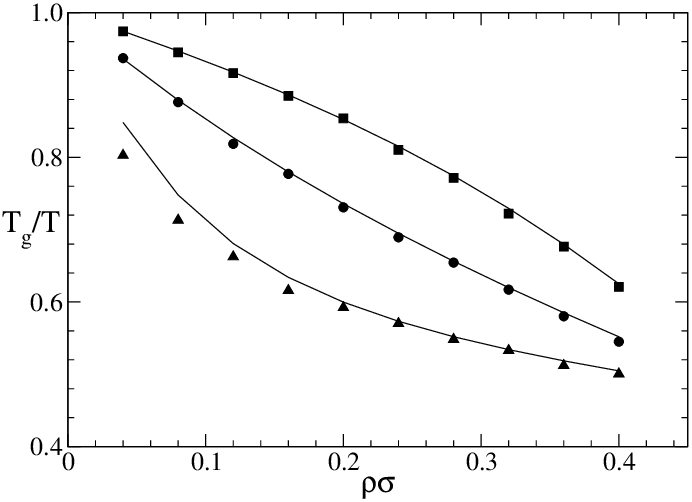}
\caption{
\label{fig:temperatura}
Dependence of the ratio between granular $T_g$ and bath temperature $T=10$ 
on the density,
for a system with repulsive $\epsilon=-10$ (squares), vanishing $\epsilon=0$ 
(circles) and attractive $\epsilon=10$ (triangles) interparticle interaction. 
Points indicate the average over a set of $10^4$ samplings 
in a single MD run of duration $t_{max}=10^4$.
Solid lines refer to theoretical temperatures extracted
from the numerical solution of Eq.~(\ref{eq:Tstat}). 
The number of particles is $N=2000$, the remaining parameters
are chosen as $\alpha=0.9$, $\gamma=0.2$ and $b=2$.}
\end{figure}

\begin{figure}[htb]
\includegraphics[clip=true,width=8.cm, keepaspectratio,angle=0]
{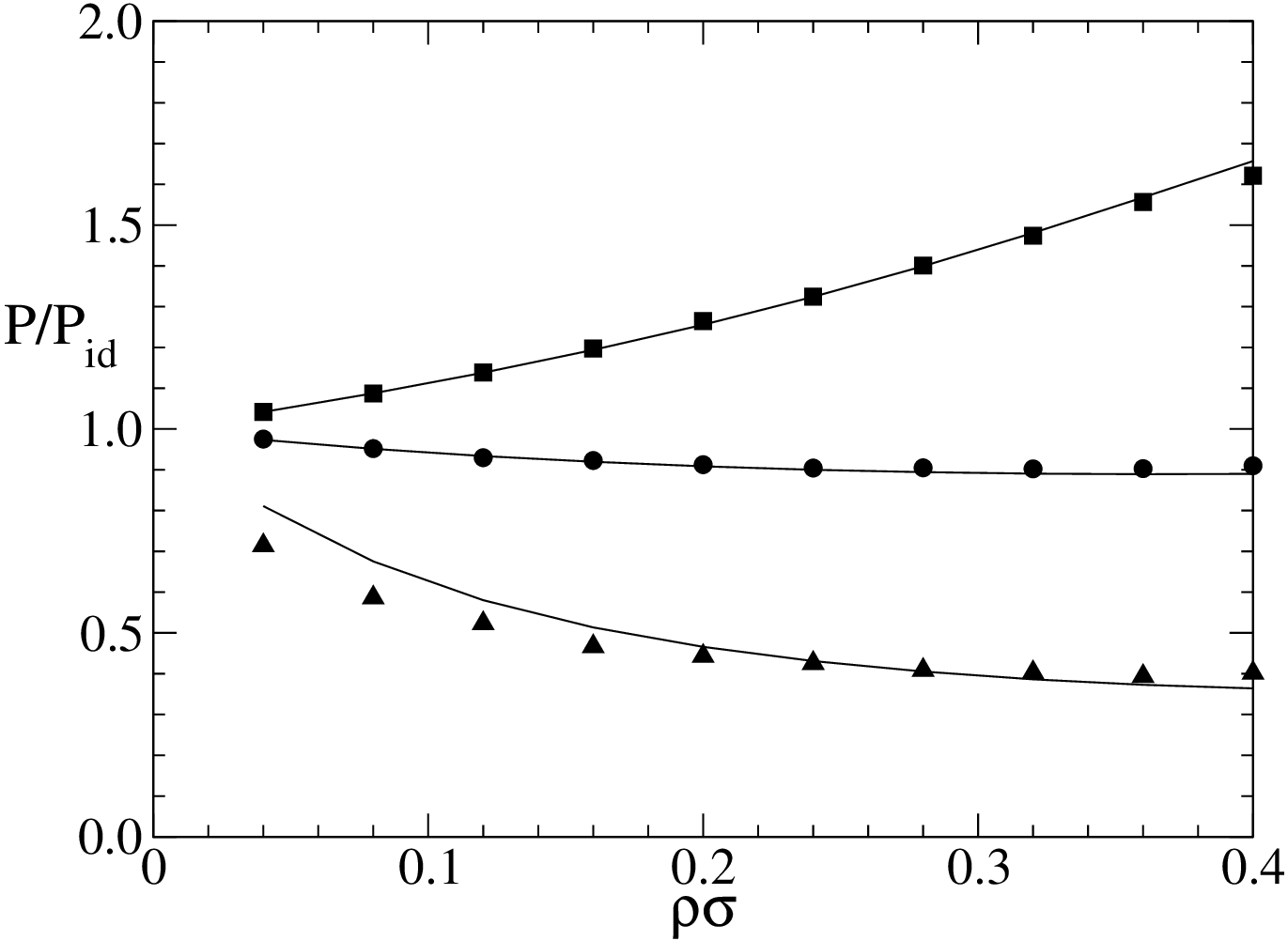}
\caption{Pressure of an inelastic system, rescaled to the equivalent 
ideal gas pressure ($P_{id} = \rho T$), as a function of $\rho\sigma$ in the 
case of $\epsilon=-10$
(squares), $\epsilon=0$ (circles) and $\epsilon=10$ (triangles), for a 
system with the same parameters as in Fig.~\ref{fig:temperatura}. 
Solid lines are the corresponding analytical values obtained according to 
formula~(\ref{veroviriale}).}
\label{fig:pressione}
\end{figure}

The scenario changes when the system is coupled to
an heat-bath at temperature $T$. A steady
regime, characterized by almost constant granular temperature
and pressure, is attained. As already done for the cooling regime, 
we can derive an implicit relation
for $T_g$ \cite{dererum1,dererum2} by multiplying both sides of 
Eq.~(\ref{evo}) by $v^2$ and integrating with respect to $v$,
\begin{equation}
\bigg[1+(1-\alpha^2)\frac{\omega_0(T_g)}{2\gamma}\bigg]\; T_g = T
\label{eq:Tstat}
\end{equation}
The variation of $T_g$ with density, given by the numerical solution of 
Eq.~(\ref{eq:Tstat}), is compared in figure~\ref{fig:temperatura} 
with the results of MD simulations. 
The agreement between theory and numerical
experiments is satisfactory for the three possible cases:
attractive, repulsive and vanishing inter-particle interaction.
The virial formula (\ref{Pg}) is employed to compute the pressure of the 
system by averaging over different MD runs. The simulated pressure values are 
plotted, in figure \ref{fig:pressione}, together with  
those obtained by a self-consistent solution of formula~(\ref{veroviriale})
with the appropriate replacement of the
heat-bath temperature $T$ by the granular temperature $T_g$, 
Eq.~(\ref{eq:Tstat}).

\begin{figure}[htb]
\includegraphics[clip=true,width=8.cm, keepaspectratio,angle=0]
{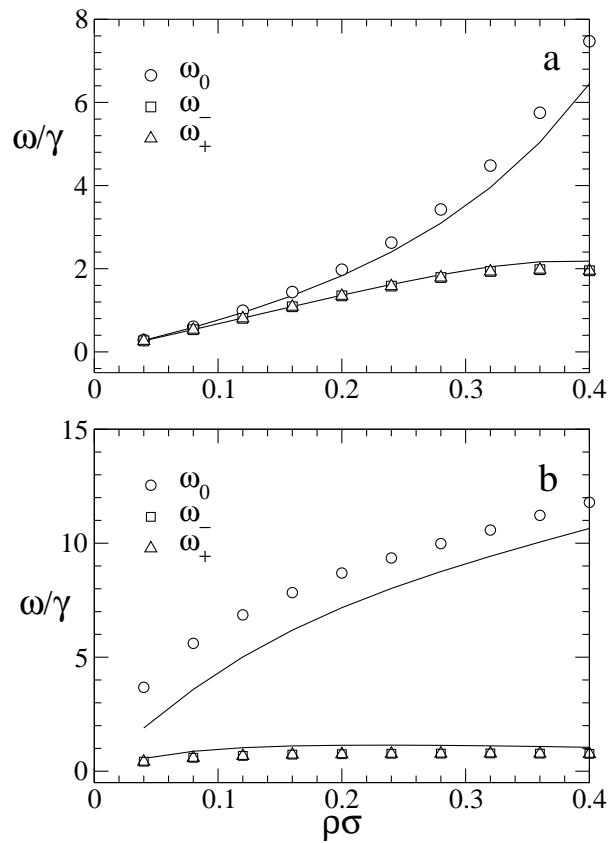}
\caption{
Collision frequencies at particle separation $x_{ij}=\sigma$ and
$x_{ij}=b\sigma$ as a function of $\rho\sigma$.
Figure (a) shows the repulsive case ($\epsilon=-10$) while
figure (b) refers to attractive interaction ($\epsilon=10$).
Lines correspond to the theory from Eqs.~(\ref{eq:frequencies}).
The remaining parameters are as in Fig.~\ref{fig:temperatura}.}
\label{fig:freq}
\end{figure}

\begin{figure}[htb]
\includegraphics[clip=true,width=8.cm,keepaspectratio,angle=0]
{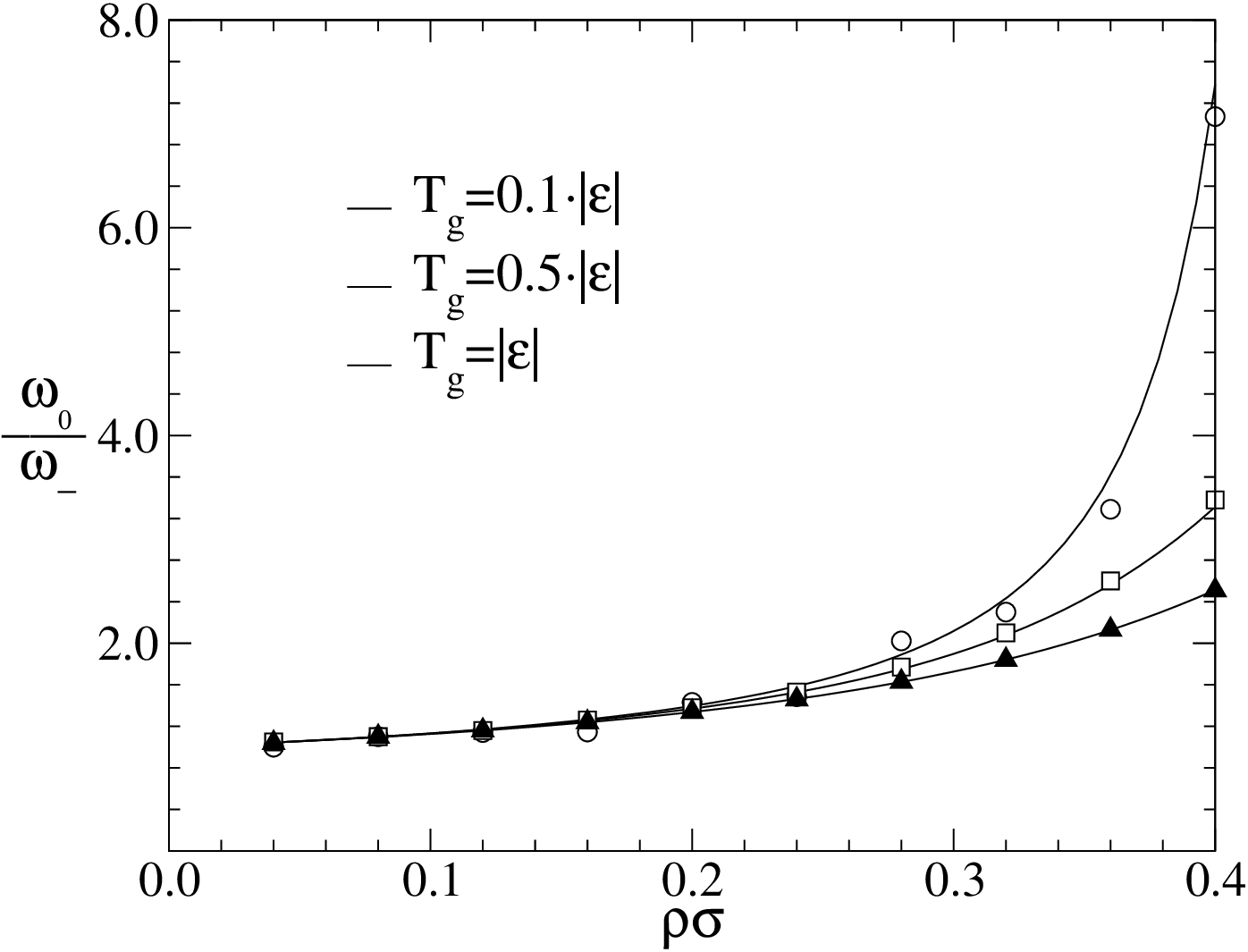}
\caption{Collision ratio (Eq.~\ref{eq:omegab}),
as a function of $\rho\sigma$ at temperatures
$T_g=|\ep|$, $T_g=5|\ep|$ and $T_g=10|\ep|$, for a driven system of
$N=2000$ particles with inelasticity
$\alpha=0.99$ and a barrier of sizes $b=2$, $\epsilon=-1$ in contact with
a bath of viscosity $\gamma=0.2$. Each point is the average over
a single run of duration $t_{max}=10^4$, sampled every $t_{sample} = 1.0$
time units = $1/\gamma$.}
\label{confr}
\end{figure}

The use of formula~(\ref{veroviriale}) implicitly assumes that 
the pair correlation function for the inelastic model maintains the same 
functional dependence as its 
equilibrium counterpart. Such an hypothesis can be checked by measuring 
during MD runs the three collision frequencies $\omega_0$, $\omega_{+}$ and
$\omega_{-}$ and comparing them with their theoretical prediction. 
The behaviour of these quantities with the dimensionless variable
$\rho \sigma$ is reported in Fig.~\ref{fig:freq}, for both attractive
and repulsive interactions.
Even in the inelastic case, one observes that
the ratio, $\omega_{0}/\omega_{\pm}$, 
between frequencies of dissipative collisions and barrier crossing
increases with density from the value $1$, observed in a very
diluted system, as shown in Fig.~\ref{confr}. This is very 
consitent with the prescription provided by formula~(\ref{eq:omegab}) 
that, hard-core collisions, in this model, become dominant 
events at higher densities.   


\begin{figure}[t]
\includegraphics[clip=true,width=16.cm, keepaspectratio,angle=0]
{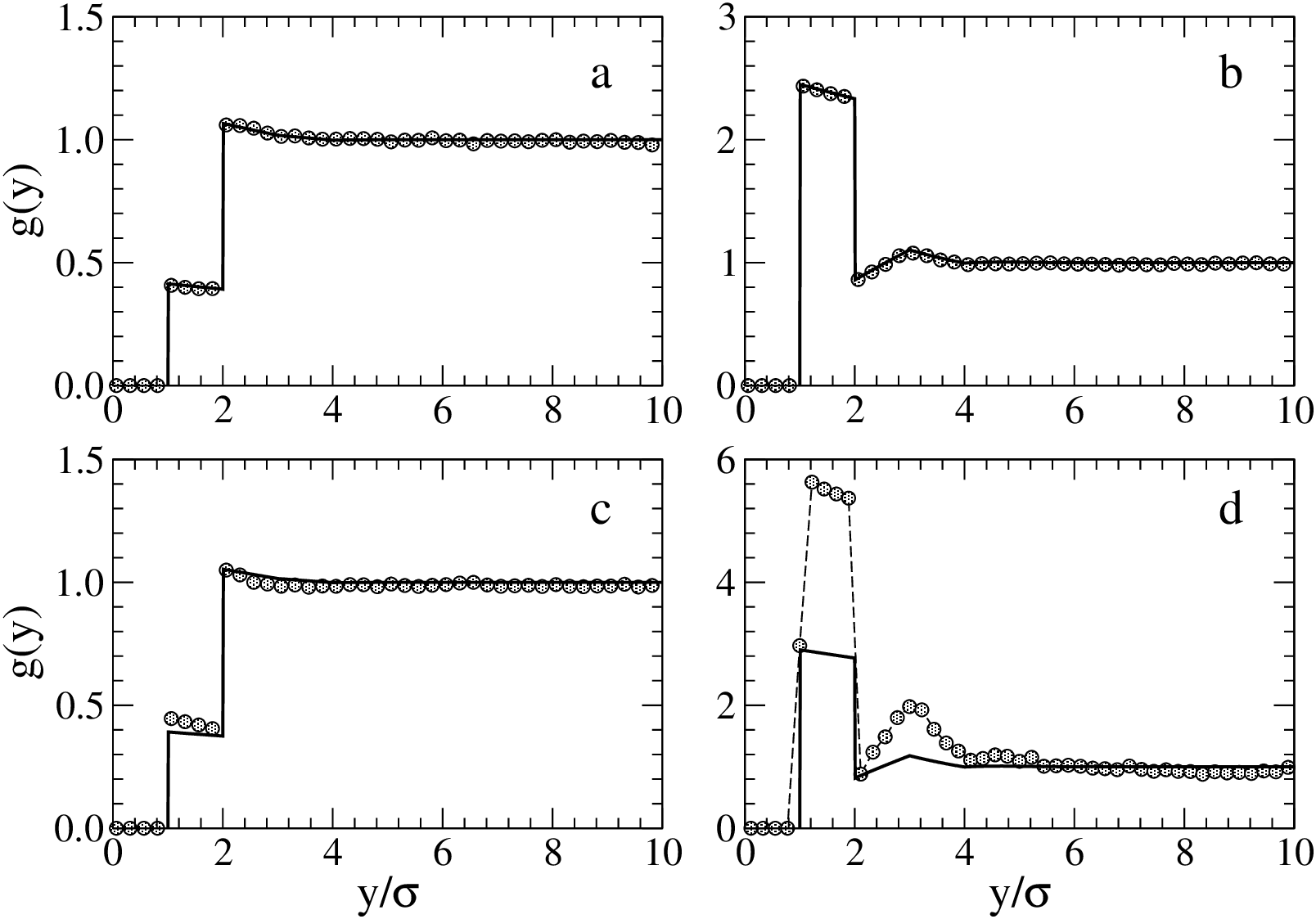}
\caption{Simulated (circles) and theoretical (solid line)
pair correlations as a function of the rescaled distance $y/\sigma$, 
for system 
of $N=2000$ hard-rods with repulsive-elastic a) attractive-elastic b),
repulsive-inelastic ($\alpha=0.9$) c),  attractive-inelastic ($\alpha=0.9$) d)
interactions, 
and parameters $T=|\ep|$, $\ep=10$, $\gamma=0.2$,  
$\rho\sigma=0.05$ and $b=2$. Points represent the average of the function
over $10^4$ sampled configurations extracted from run of length of $t=10^4$.}
\label{fig:gr}
\end{figure}

In the case of barriers (Fig.~\ref{fig:freq}a), the theoretical 
frequencies agree fairly with those extracted from simulations. 
However, some discrepancies arise when particles may mutually 
attract (Fig.\ref{fig:freq}b), even though the overall trend of the 
frequencies with the particle density is correctly \
captured by the theoretical predictions.
The differences induced by inelasticity become more evident 
in Fig.~\ref{fig:gr}, 
where we plot the theoretical and numerical pair correlation
functions $g(y)$ at the value $\rho\sigma=0.05$ for repulsive c) 
and attractive d) particles. Again, the theory fits faithfully the 
simulations for the system with repulsive interactions, while, for 
attracting particles, the simulated $g(y)$ deviates of about a factor two 
from its estimate in the interval $\sigma\le y \le b\sigma$. 
Figures (Fig.\ref{fig:gr}a) and (Fig.\ref{fig:gr}b), on the contrary, 
indicates clearly that our theoretical approach perfectly describes
the functions $g(y)$ of the elasitc system with both attractive and   
repulsive forces. 

The difficulty encountered by the theory to fit  
some regimes of the system with attracting inelastic particles can be 
ascribed to the different effects that repulsive and
attractive interactions induce on the inelastic system. 
The former basically entails a system re-elasticization which may favor 
homogeneous particle distributions,   
while the latter enhances the frequency of inelastic collisions 
leading to clustering. The relevant physical parameter controlling the 
system behaviour is the ratio $|\ep|/T_g$, thus a good evaluation of $T_g$ 
is crucial to make accurate theoretical predictions. 
Our estimate of the granular temperature $T_g$, from Eq.~(\ref{eq:Tstat}), 
is based on the assumptions of spatial homogeneity. 
For repulsive interactions (barriers) the homogeneous state occurs,
while for cohesive interactions,
particles, under specific conditions, can easily cluster making the 
system inhomogeneous.
If this happens, the single observable, $T_g$, does not describe 
properly the kinetic state of the system and, in addition, its 
estimate from Eq.~(\ref{eq:Tstat}) is incorrect 
since that formula neglects local temperature fluctuations.

The deviations of the theory from simulations become less pronounced 
as $\gamma$ increases, and the reliability of the theoretical 
approach can be quantified by the integrated difference between
numerical, $g_n(y)$, and theoretical, $g_t(y)$,   
\begin{equation}
\Delta g=\frac{1}{\sigma} \int_{\sigma}^{b\sigma} dy [g_n(y)-g_t(y)]
\label{eq:area}
\end{equation}
for various values of $\gamma$, but $\ep/T$ = const. 
The dependence of $\Delta g$ on $\gamma$ reflects the fact that the
response of the fluid to the action of the heat-bath is faster as
$\gamma \to \infty$ and thus erases more rapidly
the memory of inelastic collisions. Within this limit one recovers the
behavior of the elastic system.  
\begin{figure}[t]
\vspace{0.5cm}
\includegraphics[clip=true,width=8.0cm,keepaspectratio]
{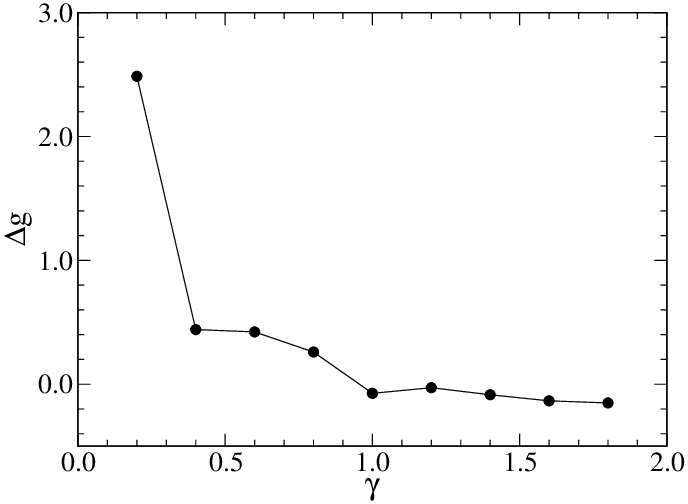}
\caption{Parametric plot of the cumulated difference $\Delta g$ between 
theoretical and simulated correlations function (see  Eq.~(\ref{eq:area})) 
versus the friction coefficient $\gamma$, for an 
inelastic system with $N=2000$ rods and parameters 
$\alpha=0.9$, $T=10$, $\epsilon=10$, $\rho\sigma=0.05$ and $b=2$.}
\label{fig:corr}
\end{figure}

\section{Conclusions}
In this paper we have investigated both theoretically and numerically the 
influence of a finite range interparticle interaction on 
the behavior of a one-dimensional inelastic hard-rod system. 
Forces and interactions, whose range is larger than the size associated 
with the excluded volume constraint are often present in many realistic 
granular materials.
In the specific case, we have chosen a square well potential to model 
attraction and a square barrier to model repulsion. 
These simple shapes, in the case of undriven system, still
enable a computer implementation of the particle evolution in terms of a 
collision driven molecular dynamics.
In fact, simple transformations describe the 
instantaneous changes of velocities when the separation 
between two particle corresponds to the two characteristic ranges
of the potential.   
We first analyzed how the interplay 
between these finite range forces and inelasticity  
modifies the cooling scenario with respect to the free inelastic system. 
We found that in the case of
repulsive barriers the temperature decay becomes slower
than Haff's $1/t^2$ power law and eventually reaches a regime 
where the system is nearly elastic. In the case of attractive
wells, instead, the granular temperature is lost faster
than an inverse time power law.

Secondly, we studied the behavior of the stationary regime obtained
through a stochastic forcing of the system. The steady state 
has been analyzed via MD simulations and theoretical 
approaches based on the direct comparison with the elastic  
counterpart of the system whose equilibrium properties are 
well understood. Our results show that, in the dense limit, particle 
spatial correlations are relevant and modify the collision rate, 
the excluded volume of the other particles
enhances the probability that two particles are at contact and thus 
reduce the repulsive barrier. 
The theoretical approach we have attempted remains a reliable approximation 
for the behaviour of the dissipative system at not too small densities while 
it is correct for the elastic system at every density. 

\section*{Acknowledgments}
U. Marini-Bettolo-Marconi acknowledges the financial support of the 
Project Complex Systems and Many-Body Problems
Cofin-MIUR 2003 prot. 2003020230.



\begin{thebibliography}{0}
\bibitem{note}
To the best of our knowledge the pair correlation function for
the one-dimensional hard-rod fluid had not been obtained previously.
\end{thebibliography}
\end{document}